\documentclass[]{cccg}  
\usepackage{graphicx}
\usepackage{url}
\usepackage{amsfonts}
\usepackage{amssymb}
\usepackage{latexsym}

\newenvironment{pf}{\unskip{\bf Proof:}}{\unskip{\hfill $\Box$}}

\newcommand{\lemlab}[1]{\label{lemma:#1}}
\newcommand{\theolab}[1]{\label{theo:#1}}

\newcommand{\eqlab}[1]{\label{eq:#1}}
\newcommand{\corlab}[1]{\label{cor:#1}}

\newcommand{\figlab}[1]{\label{fig:#1}}
\newcommand{\seclab}[1]{\label{section:#1}}

\newcommand{\lemref}[1]{\ref{lemma:#1}}

\newcommand{\theoref}[1]{\ref{theo:#1}}

\newcommand{\corref}[1]{\ref{cor:#1}}

\newcommand{\figref}[1]{\ref{fig:#1}}
\newcommand{\eqref}[1]{(\ref{eq:#1})}
\newcommand{\secref}[1]{\ref{section:#1}}

\newtheorem{theorem}{Theorem}
\newtheorem{lemma}{Lemma}
\newtheorem{cor}{Corollary}

{\catcode`\@=11
\gdef\setft#1#2#3{%
\def\@oddfoot{
{\setbox0=\hbox{#1}
\setbox1=\hbox{#3}
\ifdim\wd0>\wd1
\dimen0=\wd0
\box0\hfil#2\hfil\hbox to\dimen0{\hfil\hfil\box1}
\else \dimen0=\wd1
\hbox to\dimen0{\box0\hfil }\hfil#2\hfil\box1 \fi
}}} }

\def\complaint#1{}
\def\withcomplaints{
\newcounter{mycomplaints}
\def\complaint##1{\refstepcounter{mycomplaints}%
\ifhmode%
\unskip%
{\dimen1=\baselineskip \divide\dimen1 by 2 %
\raise\dimen1\llap{\tiny -\themycomplaints-}}\fi%
\marginpar{\tiny [\themycomplaints]: ##1}}%
}

\let\oldendpf=\endpf
\def\endpf{\oldendpf\par\medskip}

\def\a{{\alpha}}
\def\b{{\beta}}
\def\g{{\gamma}}
\def\G{{\Gamma}}
\def\l{{\lambda}}
\def\s{{\sigma}}
\def\C{{\cal C}}

\def\t{{\theta}}
\def\p{{\phi}}

\begin{document}

\title{{\bf On the Development of the Intersection\\
of a Plane with a Polytope}}

\author{%
Joseph~O'Rourke%
\thanks{
Dept.\ Comput.\ Sci., Smith Col\-lege, North\-ampton,
MA 01063, USA.
\texttt{orourke@\penalty \exhyphenpenalty cs.smith.edu}.
Supported by NSF grant CCR-9731804.}
}

\date{}

\maketitle

\markboth{}{}

\begin{abstract} 
Define a ``slice'' curve as the intersection of
a plane with the surface of a polytope, i.e., a convex
polyhedron in three dimensions.  We prove that a slice curve
develops on a plane without self-intersection.
The key tool used is a
generalization of
Cauchy's arm lemma to permit nonconvex ``openings'' of
a planar convex chain.
\end{abstract}
 
\section{Introduction}
\seclab{Introduction}
Although the intersection of a plane $\Pi$ with a polytope $P$
is a convex polygon $Q$ within that plane, on the surface
of $P$, this ``slice curve'' can be nonconvex, alternatively turning
left and right.  The development of a curve on a plane is
determined by its turning behavior on the surface.  Thus slice curves
develop (in general) to nonconvex, open chains on a plane.
The main result of this paper is that slice curves always
develop to simple curves, i.e., they do not self-intersect.

Our main tool is a generalization of an important lemma
Cauchy used to prove the rigidity of polytopes.
Cauchy's arm lemma says that if $n-2$ consecutive angles
of a convex polygon are opened but not beyond $\pi$, 
keeping all but one edge length
fixed and permitting that ``missing'' edge $e$ to vary
in length, then
$e$ lengthens (or retains its original length).
We employ a generalization of this lemma to permit opening of the
angles beyond $\pi$, as far reflex as they were originally convex.
The conclusion remains the same: $e$ cannot shorten.
We will see that this conclusion follows from a theorem of
Axel Schur~\cite{s-usedkkk-21}.

The first part of this paper (Section~\secref{Cauchy})
concentrates on this generalization
of Cauchy's lemma.  The issue of self-intersection is addressed
in Section~\secref{Noncrossing}, and the curve development
result is proved in Section~\secref{slice}.

\section{Cauchy's Arm Lemma Extended}
\seclab{Cauchy}
Let $A = (a_0,a_1,\ldots,a_n)$ be an $n$-link polygonal chain
in the plane
with $n$ fixed edge lengths $\ell_i = |a_i a_{i+1}|$, $i=0,\ldots,n-1$.
We call the vertices $a_i$ the {\em joints\/} of
the chain, $a_0$ (which will always be placed at the origin)
the {\em shoulder\/}, and $a_n$ the {\em hand}.
Define the {\em turn angle\/} $\a_i$ at joint $a_i$, $i=1,\ldots,n-1$
to be the angle in $[-\pi,\pi]$ that turns the vector
$a_i - a_{i-1}$ to $a_{i+1} - a_i$, positive for left (counterclockwise)
and negative for right (clockwise) turns.

Define an open polygonal chain $A$ to be {\em convex\/} if its joints
determine a (nondegenerate) convex polygon, i.e.,  all joints are distinct
points (in particular, $a_n \neq a_0$),
all joints lie
on the convex hull of $A$ and they do not all lie on a line.
Note there is no chain link between $a_n$ and $a_0$.
The turn angles for a convex chain all lie in $[0,\pi)$; but note
this is not a sufficient condition for a chain to be convex, for
it is also necessary that the angles at $a_0$ and $a_n$ be
convex.

We can view the configuration of a polygonal chain $A$ to
be determined 
by two vectors:  the fixed edge lengths
$L = (\ell_0,\ldots,\ell_{n-1})$
and the variable turn angles $\a = (\a_1,\ldots,\a_{n-1})$,
with the convention that $a_0$ is placed at the origin
and $a_n$ horizontally left of $a_0$.
Let $\C_L(\a) = A$ be the configuration so determined.
We use $\a$ to represent the angles of the initial configuration,
and $\b$ and $\g$ to represent angles in a reconfiguration.

Let $D(r) = \{p \;:\; |p a_0| < r \}$ be the open disk of radius $r$
centered on the shoulder joint $a_0$.
Define $a = |a_n a_0|$, the length of 
the ``missing'' link, the original hand-to-shoulder
distance.  Finally, we will call $D(a)$ the {\em forbidden (shoulder) disk}.
We may state Cauchy's arm lemma in the following form:

\begin{theorem}
If $A = \C_L(\a)$ is a convex chain with fixed edge lengths $L$, and
turn angles $\a$, then in any reconfiguration to
$B = \C_L(\b)$ with turn angles $\b = (\b_1,\ldots,\b_{n-1})$
satisfying
\begin{equation}
\b_i \in [ 0, \a_i ]
\eqlab{convex}
\end{equation}
we must have $|b_n b_0| \ge |a_n a_0|$,
i.e., the hand cannot enter the forbidden disk $D(a)$.
\theolab{cauchy}
\end{theorem}

Cauchy's lemma is sometimes known as Steinitz's lemma,
because Steinitz noticed and corrected an error in the
proof a century after Cauchy~\cite[p.~235]{c-p-97}.
Many proofs of Cauchy's lemma are now known, e.g.,
\cite{sz-clccp-67,s-gpf-97}
and
\cite[p.~64]{az-pftb-98}.

Our main tool is a generalization of Cauchy's lemma
that replaces the $0$ in Eq.~\eqref{convex} by $-\a_i$,
and is otherwise identical:

\begin{theorem}
If $A = \C_L(\a)$ is a convex chain with fixed edge lengths $L$, and
turn angles $\a$, then in any reconfiguration to
$B = \C_L(\b)$ with turn angles $\b = (\b_1,\ldots,\b_{n-1})$
satisfying
\begin{equation}
\b_i \in [ -\a_i, \a_i ]
\eqlab{turn.range}
\end{equation}
we must have $|b_n b_0| \ge |a_n a_0|$,
i.e., the hand cannot enter the forbidden disk $D(a)$.
\theolab{cgen}
\end{theorem}
The intuition is illustrated in 
Fig.~\figref{arm1}; further examples
are provided in Fig.~\figref{examples}.
\begin{figure}[htbp]
\centering
\includegraphics[width=0.9\linewidth]{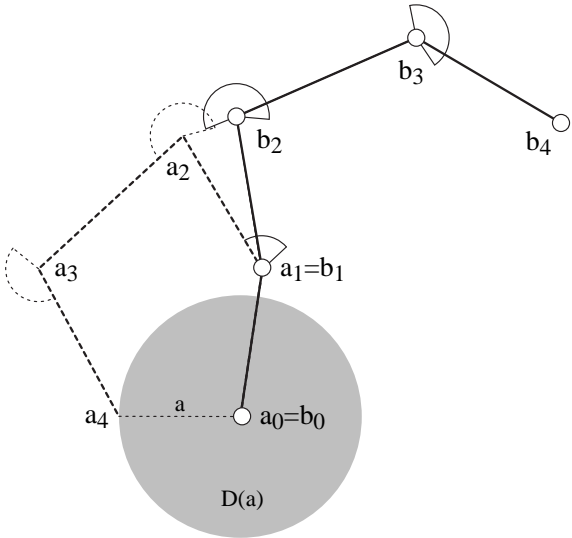}
\caption{A reconfiguration of a $4$-link convex chain 
$A$ to chain $B$, satisfying
Eq.~\eqref{turn.range}, leaves $b_n$ outside $D(a)$.
The valid angle ranges are marked by circular arcs.}
\figlab{arm1}
\end{figure}
Although the chain may become nonconvex,
Eq.~\eqref{turn.range} ensures that the movement constitutes
a form of straightening.
Note that Theorem~\theoref{cgen} makes no claim
about steadily increasing hand-shoulder separation during
some continuous movement to $B$; indeed a continuous
opening could first increase and later decrease the separation.
Rather the claim is that a final configuration satisfying
Eq.~\eqref{turn.range} cannot place the hand in the forbidden disk.

As pointed out by Connelly in~\cite[p.~30]{c-re-82},
Schur generalized Cauchy's theorem to the smooth case~\cite{s-usedkkk-21}.
Rather than requiring angles to open, Schur requires the curvature
to decrease. Some mentions of Schur's theorem in the
literature, e.g., in~\cite[p.31]{g-dg-63}, phrase it as the smooth,
planar equivalent of Cauchy's lemma, which does not
capture the nonconvexity permitted in the statement of Theorem~\theoref{cgen}.
But others, notably the exposition by Chern~\cite[p.~119]{c-cses-89},
state it as a generalization also to space curves,
and employ the absolute value of curvature,
implicitly permitting nonconvexity.
Chern also states it (without proof) in the piecewise-smooth case,
which directly encompasses polygonal chains.
Although it remains unclear whether these authors
intended to capture the precise statement of Theorem~\theoref{cgen},
we will show that Chern's version of Schur's proof
can be used to establish the theorem.
Consequently, Theorem~\theoref{cgen} can be seen as a consequence
of Schur's theorem, if not a direct corollary to it.

As with Cauchy's arm lemma, one may expect many different proofs
of such a fundamental result.
We offer three proofs of Theorem~\theoref{cgen} in this paper.
The first and second are induction proofs, following the same
general outline.
The third follows Chern's proof of Schur's theorem, specializing
the smooth, space-curve argument to the nonsmooth, planar situation.
The reader uninterested in these proofs may skip to
Section~\secref{Noncrossing}.

\subsection{First Proof of  Theorem~\theoref{cgen}}
\seclab{proof.1}

Although we impose no restriction on self-intersection of the
chain, we will show in Theorem~\theoref{noncrossing} that
the chain remains simple.
Note that, because we fix $a_0$ to the origin, and the first turn
angle is at joint $a_1$, in any reconfiguration the first edge of the
chain is fixed.

Our first proof of Theorem~\theoref{cgen}
requires a few preparatory lemmas.
We start with the simple observation that negating the turn angles reflects
the chain.
\begin{lemma}
If a chain $A = \C_L(\a)$ is reconfigured to
$B = \C_L(\b)$
with
$\b_i = -\a_i$, $i=1,\ldots,n-1$,
then $B$ is a reflection of $A$ through the line $M$ containing $a_0 a_1$,
and $|b_n b_0| = |a_n a_0|$.
\lemlab{reflect}
\end{lemma}
\begin{pf}
Reflecting $A$ through line $M$ does indeed negate each turn
angle: $\b_1 = -\a_1$ is immediate, and all others have
their sense reversed by the reflection.  Because $\b$ determines
the configuration uniquely, this reflection is indeed the
configuration determined by that $\b$ vector.
Because $M$ passes through $a_0=b_0$, $b_n$ remains at the same
distance from $b_0$ as $a_n$ is from $a_0$.
\end{pf}

Call a reconfiguration $B=\C_L(\b)$ of a convex chain $A = \C_L(\a)$
which satisfies the constraints of Eq.~\eqref{turn.range}
a {\em valid reconfiguration},
and call the vector of angles $\b$ {\em valid angles}.
Define the {\em reachable region\/} $R_L(\a)$
for a convex chain $A = \C_L(\a)$ 
to be the set of all hand positions $b_n$ for any valid reconfiguration
$B = \C_L(\b)$.
One can view Theorem~\theoref{cgen} as the claim that
$R_L(\a) \cap D(a) = \emptyset$.
It is well known~\cite{hjw-mp2dl-84}\cite[p.~326]{o-cgc-98} 
that the reachable region for a chain with no
angle constraints is a shoulder-centered closed annulus,
but angle-constrained reachable regions seem unstudied.

For the first proof 
we need two technical lemmas.

\begin{lemma}
The configuration of a chain $A = \C_L(\a)$ is a continuous function
of its turn angles $\a$.
\lemlab{continuous}
\end{lemma}
\begin{pf}
The coordinates of each joint $a_i$ can be written as a
trigonometric polynomial (rotation and translation
of each link), with terms multiplying
$\sin()$ and $\cos()$
applied to angles, and constants depending on the lengths $L$.
Since all the constituents of these polynomials are continuous
functions of the angles, each joint, and so all joints, are
also.
\end{pf}

\begin{lemma}
$R_L(\a)$ is a closed set.
\lemlab{closed}
\end{lemma}
\begin{pf}
The $(2n{-}2)$-dimensional configuration space $S$ of all chains
$B=(b_0,b_1,\ldots,b_n)$ with valid $\b=(\b_1,\ldots,b_{n-1})$
is the image of
the trigonometric polynomials mentioned in the previous
proof as the angles
vary over the compact
domain
$$
[-\a_1,\a_1] \times \cdots \times [-\a_{n-1},a_{n-1}]
$$
Because the image of a continuous function on a compact
domain is compact, and because the function is continuous
by Lemma~\lemref{continuous},
$S$ is compact.
In Euclidean space, a compact set is closed and bounded;
so $S$ is closed.
$R_L(\a)$ is just the $2$-dimensional $b_n$-slice through $S$,
and so it is closed as well.
\end{pf}

We use this lemma to help identify, among potential counterexamples,
the ``worst'' violators.
Define a configuration $B = \C_L(\b)$ to be {\em locally minimal\/}
if there is a neighborhood $N$ of $\b$ 
such that, for all $\b' \in N$, the determined hand position
$b'_n$ is no closer to the shoulder:
$|b'_n a_0| \ge |b_n a_0|$.
Thus the hand's distance to the shoulder is locally minimal.

\begin{lemma}
Let $B = \C_L(\b)$ be a reconfiguration of convex chain $A = \C_L(\a)$
with $b_n \in D(a)$.  Then either $b_n = a_0$, or
there is some locally minimal configuration $B' = \C_L(\b')$
with $b'_n \in D(a)$.
\lemlab{loc.min}
\end{lemma}
\begin{pf}
Suppose $b_n \neq a_0$.  Inflate a circle $C(r)$ about $a_0$,
starting with radius $r=0$, until some point of $R_L(\a)$
is first encountered.  
Because $b_n \in D(a)$, this event will occur before 
$r=a$.
Because $R_L(\a)$ is closed by Lemma~\lemref{closed},
there is some definite, smallest radius $r_0$, $0 < r_0 < a$,
when the circle first hits the reachability region.
A configuration corresponding to any point in
$C(r_0) \cap R_L(\a)$ satisfies the lemma.
\end{pf}

\noindent
The above lemma will provide a ``hook'' to reduce $n$ in the induction step.
We separate out the base of the induction in the next lemma.

\begin{lemma}
Theorem~\theoref{cgen} holds for $n=2$.
\lemlab{base}
\end{lemma}
\begin{pf}
A $2$-link chain's configuration is determined by single angle at $a_1$.
The reachable region is a single circular arc exterior to $D(a)$,
centered on $a_1$, of radius $\ell_1$.
See Fig.~\figref{base}.
\end{pf}
\begin{figure}[htbp]
\centering
\includegraphics[width=0.8\linewidth]{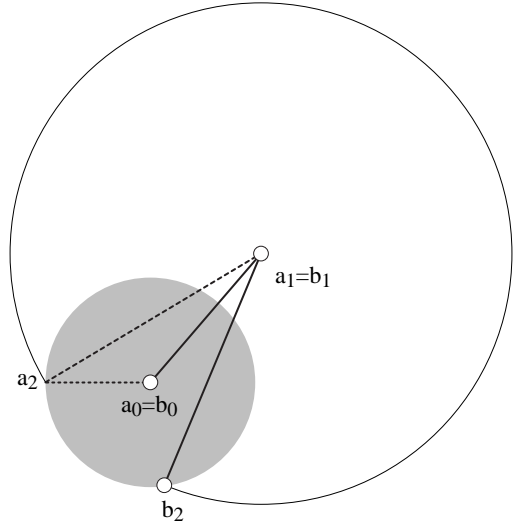}
\caption{$R_L(\a)$ for a $2$-link chain is a circle arc centered on $a_1=b_1$.}
\figlab{base}
\end{figure}

\noindent
We now prove Theorem~\theoref{cgen} by induction.

\noindent
\begin{pf}
Lemma~\lemref{base} establishes the theorem for $n=2$.
Assume then that the theorem holds for all chains of $n-1$ or
fewer links.  We seek to establish it for an $n$-link
chain $A = \C_L(\a)$, $n > 2$.
Assume, for the purposes of contradiction, that $A$ may
be reconfigured so that the hand falls inside the forbidden
disk $D(a)$.  We seek a contradiction on a shorter chain.
By Lemma~\lemref{loc.min}, one of two cases
holds:  the hand reaches $a_0$, or there is a locally minimal
configuration.

\begin{enumerate}
\item
Suppose $B=\C_L(\b)$ is such that $b_n=a_0$,
as illustrated in
Fig.~\figref{case1}(c).
There are two possibilities.
Either $\ell_{n-1}= |a_{n-1}a_n| < |a_{n-1}a_0| = a'$,
when $a_{n-1}$ is left of the bisector of $a_0 a_n$
(Fig.~\figref{case1}(a)),
or $a_{n-1}$ is right of the bisector
(Fig.~\figref{case1}(b)).
In the latter case, because $a_1$ cannot be left
of the bisector, if we relabel the chain in reverse,
then we again have $\ell_{n-1} < a'$.
(Note that if both $a_{n-1}$ and $a_1$ are on the bisector,
then $a_{n-1}=a_1$ and the chain has only two links.)

Now consider the chains $A'$ and $B'$ obtained by removing
the last links $a_{n-1} a_n$ and $b_{n-1} b_n$.
First, $A'$ is a convex chain of $n-1$ links, so the
induction hypothesis applies and says that $A'$ cannot
be validly reconfigured to place $b_{n-1}$ closer to $a_0$
than $a'=|a_{n-1} a_0|$.  $B'$ places $b_{n-1}$ at
distance $\ell_{n-1}$ from $a_0$, which we just
observed is less than $a'$.
It remains to argue that $B'$ is a valid reconfiguration of
$A'$, i.e., that it satisfies Eq.~\eqref{turn.range}.
However, this is satisfied for $i=1,\ldots,n-2$ because
these angles are not changed by the shortening, and
after shortening there is no constraint on $\b_{n-1}$.
Thus $B'$ is a valid reconfiguration of $A'$ but places the
hand in the forbidden disk, a contradiction.
\begin{figure}[htbp]
\centering
\includegraphics[width=\linewidth]{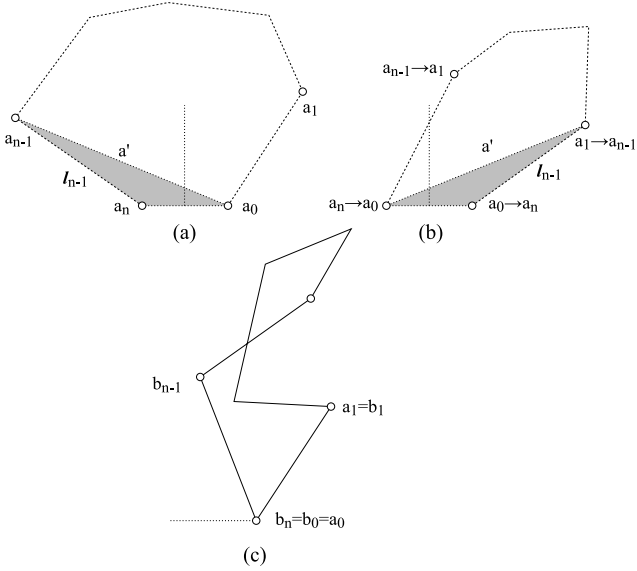}
\caption{Case 1: $b_n=a_0$. (Drawing (c) is not accurate.)}
\figlab{case1}
\end{figure}

\item  We may henceforth assume, by Lemma~\lemref{loc.min}, that there is a
locally minimal configuration $B=\C_L(\b)$ that places $b_n \in D(a)$.
Again we seek to shorten the chain and obtain a contradiction.

First we establish that at least one\footnote{
	In fact I believe that all must be extreme, but the
	proof only needs one.
}
$\b_i$ is at the limit of its valid turn
range: $\b_i= \pm \a_i$.  Suppose to the contrary that all $\b_i$,
$i=1,\ldots,n-1$, are strictly interior to their allowable turn
ranges: $\b_i \in (-\a_i,\a_i)$.
Let $M$ be the line containing $b_0 b_n$.
Consider two cases:
\begin{enumerate}
\item Some $b_i$, $i=1,\ldots,n-1$, 
does not lie on $M$.  Then because $\b_i$ is not
extreme, the subchain $(b_{i+1},\ldots,b_n)$ may be rotated
about $b_i$ in both directions.  Because $b_i$ is off $M$,
one direction or the other must bring $b_n$ closer to $b_0$,
contradicting the fact that $b_n$ is locally minimal.
\item All $b_i$ lie on $M$.  Then there must be some $b_i$ which
is extreme on $M$.  For this $b_i$, $\b_i = \pm \pi$.
But $\a_i \in [0,\pi)$: the nondegeneracy assumption bounds $\a_i$
away from $\pi$, and so bounds $\b_i$ away from $\pm \pi$.
\end{enumerate}
Henceforth let $b_i$ be a joint whose angle $\b_i$ is extreme.
If $\b_i = -\a_i$, then reflect $B$ about $b_0 b_n$ so
that $\b_i = \a_i$ is convex.
By Lemma~\lemref{reflect}, this does not change the distance from
$b_n$ to the shoulder, so we still have $b_n \in D(a)$.

We are now prepared to shorten the chains.  
Let  $A'$ and $B'$ be the chains resulting from removing $a_i$ and $b_i$
from $A$ and $B$ respectively:
\begin{eqnarray}
A' & = & (a_0,a_1,\ldots,a_{i-1},a_{i+1},\ldots,a_n) \\
B' & = & (b_0,b_1,\ldots,b_{i-1},b_{i+1},\ldots,b_n)
\end{eqnarray}
A crucial point to notice is that 
$|b_{i-1} b_{i+1}| = |a_{i-1} a_{i+1}|$ because $\b_i = \a_i$;
this was the reason for focusing on an extreme $\b_i$.
Therefore $B'$ is a reconfiguration of $A'$.
Of course both $A'$ and $B'$ contain $n-1$ links,
so the induction hypothesis applies.  
Moreover, because $i \le n-1$, the
$b_i$ removed does not affect the position of $b_n$.
So $b_n \in D(a)$ by hypothesis.
To derive a contradiction,
it only remains to show that $B'$ is a valid reconfiguration of $A'$,
i.e., one that satisfies the turn constraints~\eqref{turn.range}.

Let $\a'_{i+1}$ be the turn angle at $a_{i+1}$ in $A'$.
We analyze this turn angle in detail, and argue later that
the situation is analogous at $a_{i-1}$.
Let $\t$ be the angle of the triangle 
$\triangle_i=\triangle (a_i,a_{i+1},a_{i-1})$
at $a_{i+1}$; see
Fig.~\figref{case2}(a).
\begin{figure}[htbp]
\centering
\includegraphics[width=1.05\linewidth]{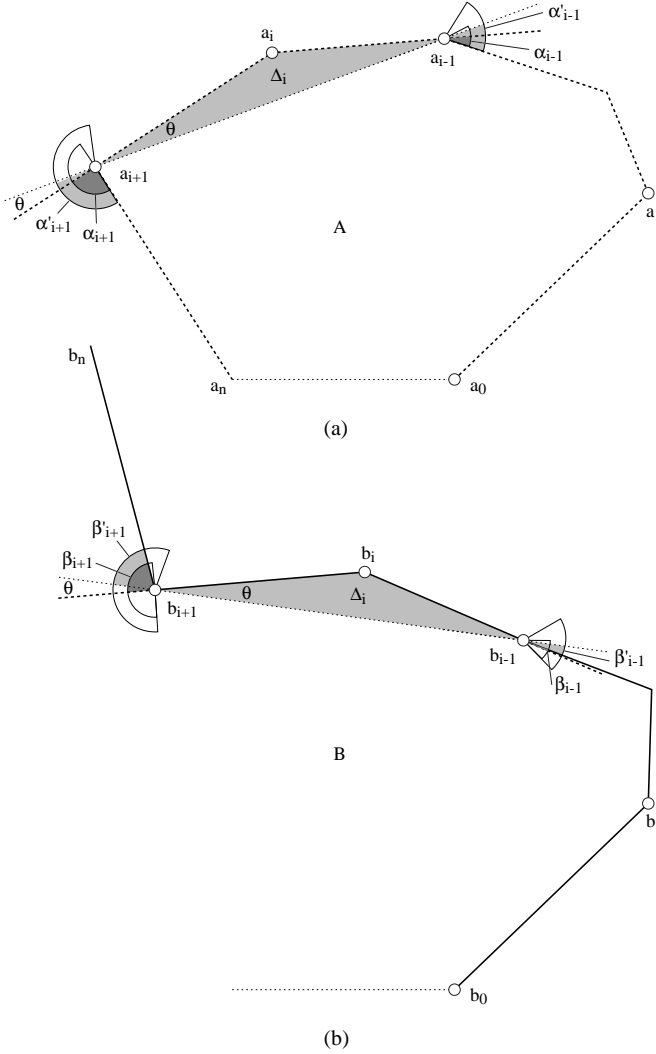}
\caption{(a) Shortening the chain $A$ by removal of $a_i$
determines new, larger turn angles $\a'_{i+1}$ and $\a'_{i-1}$
at $a_{i+1}$ and $a_{i-1}$ respectively.
(b) Here the turn angles $\b_{i+1}$ and $\b'_{i+1}$ are negative.}
\figlab{case2}
\end{figure}
Because $A$ is a convex chain,
cutting off $\triangle_i$ from $A$ increases the turn angle at $a_{i+1}$
in $A'$:
\begin{equation}
\a'_{i+1} = \t + \a_{i+1}
\eqlab{alpha}
\end{equation}

Now consider the turn angle $\b'_{i+1}$ at $b_{i+1}$ in $B'$.
Although here the turn could be negative,
as in Fig.~\figref{case2}(b),
it is still the case that the turn is advanced by $\t$
by the removal of $\triangle_i$:
\begin{equation}
\b'_{i+1} = \t + \b_{i+1}
\eqlab{beta}
\end{equation}
We seek to prove that $\b'_{i+1} \in [-\a'_{i+1},\a'_{i+1}]$.
Substituting the expressions from Eqs.~\eqref{alpha} and~\eqref{beta}
into the desired inequality yields:
$$
\begin{array}{ccccc}
-\a'_{i+1} & \le & \b'_{i+1} & \le  & \a'_{i+1} \\
-\a_{i+1} -\t & \le & \b_{i+1} + \t  & \le  & \a_{i+1} + \t \\
-\a_{i+1} -2\t & \le & \b_{i+1}  & \le & \a_{i+1}
\end{array}
$$
And this holds because $\t > 0$ and
$\b_{i+1} \in [-\a_{i+1},\a_{i+1}]$
(because $B$ is a valid reconfiguration of $A$).
The intuition here is that 
the nesting of the turn angle
ranges at
$a_{i+1}$ in $A$ and $A'$
(evident in Fig.~\figref{case2}(a))
carries over, rigidly attached to $\triangle_i$, to $B$,
so that
satisfying the tighter constraint 
in $B$ also satisfies the looser
constraint in $B'$.

Although the situation is superficially different at $a_{i-1}$
because our definition of turn angle depends on the orientation
of the chain, it is easily seen that the turn constraint is
identical if the orientation is reversed.  
Another way to view this is that we can base the turn angles
on $\triangle_i$.
Thus the equations derived
above hold again, except with $i+1$ replaced by $i-1$, and $\t$
replaced by the angle of $\triangle_i$ at $a_{i-1}$.

We have thus established that $B'$ is a valid reconfiguration of $A'$.
By the induction hypothesis, the hand $b_n$ of $B'$ cannot enter
the forbidden disk $D(a)$.  But by assumption it is in that disk.
This contradiction shows that our assumption 
that $b_n \in D(a)$ cannot hold, and establishes the theorem.
\end{enumerate}
\end{pf}

The following corollary extends the distance inequality to 
every point of the chain.
\begin{cor}
Let $A = \C_L(\a)$ be a convex chain as in Theorem~\theoref{cgen},
and let $p_1,\,p_2 \in A$ be any two distinct points of the chain.
Then in any valid reconfiguration $B$,
the points $q_1,\,q_2 \in B$ corresponding to $p_1$ and $p_2$ satisfy
$|q_1 q_2| \ge |p_1 p_2|$,
i.e., they have not moved closer to one another.
\corlab{all.points}
\end{cor}
\begin{pf}
Without loss of generality, assume that $p_1$ occurs before
$p_2$ on $A$, i.e., is a shorter length along $A$ from $a_0$.
Let  $A'$ be the chain $A$ with the portion prior to $p_1$,
and the portion after $p_2$, removed:
$A' = (p_1,\ldots,a_i,\ldots,p_2)$.
Let $B'$ be the corresponding clipped version of $B$:
$B' = (q_1,\ldots,q_2)$.
$A'$ is a convex chain because $A$ is.
$B'$ is a valid reconfiguration of $A'$, for none of the
angle ranges satisfied by $B$ have been altered.
Applying Theorem~\theoref{cgen} with $p_1$ and $q_1$ playing the
role of the shoulder, and
$p_2$ and $q_2$ the role of the hand, establishes the claim.
\end{pf}

\subsection{Second Proof of Theorem~\theoref{cgen}}
\seclab{proof.2}
We now sketch a second proof, which avoids reliance on
locally minimal configurations.
The proof is again inductive, by contradiction from a shortened
chain, and relies on the same detailed
argument concerning the turn angle ranges.
None of those details will be repeated.

\noindent
\begin{pf}
Let $A = \C_L(\a)$ be the given convex chain,
and $C = \C_L(\g)$ a valid reconfiguration that places
$c_n \in D(a)$, in contradiction to the theorem.
We first construct an ``intermediate'' configuration
$B = \C_L(\b)$ with $\b_i = |\g_i|$ for all $i=1,\ldots,n-1$,
i.e., $B$ is a convex chain formed by flipping all
turns in $C$ to be positive.
Note that, because $\g$ is a valid angle vector for $A$,
$\g_i \in [-\a_i,\a_i]$, and so $\b_i \in [0,\a_i]$.
As this is exactly the Cauchy arm opening condition, Eq.~\eqref{convex},
we may apply Theorem~\theoref{cauchy} to conclude that
$b=|b_n b_0| \ge |a_n a_0|=a$.

Now consider chain $B$.  It may be a convex chain, but it is possible 
that it is not, as in Fig.~\figref{proof.2.fix}.
\begin{figure}[htbp]
\centering
\includegraphics[width=0.8\linewidth]{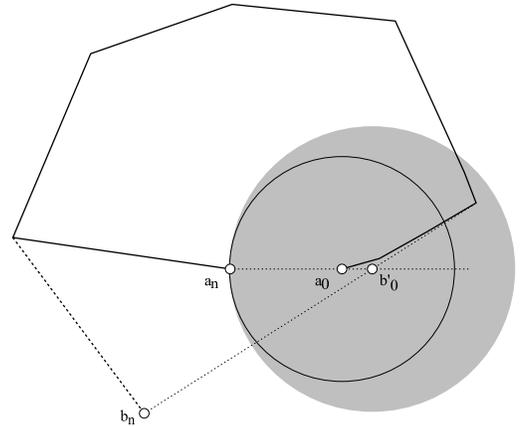}
\caption{$B$ might not be a convex chain, but another chain
starting at $b'_0$ (lying on the line through $a_0 a_n$) is.
}
\figlab{proof.2.fix}
\end{figure}
In this latter case we replace $B$ with the chain starting
at $b'_0$, which by construction is convex, and whose
forbidden disk is a superset of the forbidden disk for $A$.
To keep this sketch short, we do not further analyze the
modifications necessary in this case.

Now we focus attention on chains $B$ and $C$.
Because $\g_i = \pm \b_i$, $\g_i \in [-\b_i,\b_i]$.
Therefore, $C$ is a valid reconfiguration of $B$.
But here is the point: every angle $\g_i$ of $C$ is extreme with
respect to $B$, and
so there is no need to invoke local minimality.

Choose an $i$ and remove $b_i$ from $B$ and $c_i$ from $C$,
obtaining shorter chains $B'$ and $C'$.  Applying the
argument from the previous section verbatim, we conclude that
$C'$ is a valid reconfiguration of $B'$.
But because $B'$ has $n-1$ links, the induction hypothesis
applies and shows that $c_n$ cannot enter the forbidden
disk $D(b)$, with $b=|b_n b_0|$.  Because $b \ge a$,
$c_n$ cannot be in $D(a)$ either.  This contradicts the
assumption and establishes the theorem.
\end{pf}

\subsection{Third Proof of Theorem~\theoref{cgen}}
\seclab{proof.3}
We follow Chern's proof~\cite{c-cses-89}
of Schur's Theorem,
at times quoting Chern nearly word-for-word.
Although in some ways the specialization of his proof to
our planar, nonsmooth instance destroys some of its
elegance, the exercise does establish that
Theorem~\theoref{cgen} follows in spirit if not in
fact from Schur's Theorem.

We adjust notation slightly to more closely track Chern's
proof.
In particular, we will use $^*$ to label quantities of
the reconfigured chain.  The given chain is $A$,
and the reconfigured chain $A^* = B$ is some valid reconfiguration.

\noindent
\begin{pf}
Chern/Schur's proof is carried out largely in the domain
of the ``tangent indicatrixes'' of $A$ and $A^*$.
For any oriented curve $C$,
draw unit-length vectors parallel to the tangent vectors
of $C$, with the vectors based at an origin $O$.
Their end-points describe a curve $\G$ on the unit sphere
centered on $O$
called the {\em tangent indicatrix\/} of $C$. 
If $C$ is a curve in $d$ dimensions, the tangent indicatrix
lies on a $(d{-}1)$-dimensional sphere.
For planar curves, the indicatrix lies on a unit circle.
For polygonal curves, we turn the tangent continuously
at a vertex through the turn angle there, thus resulting
in a connected indicatrix.  The indicatrix may be ``layered''
if it turns back over itself.
An example ($\G^*$) is shown in Fig.~\figref{Schur}.
Because $A$ is a convex chain, $\G$ is a simple arc.
But $A^*$ is nonconvex and $\G^*$ doubles back over itself.
\begin{figure*}[htbp]
\centering
\includegraphics[width=0.8\linewidth]{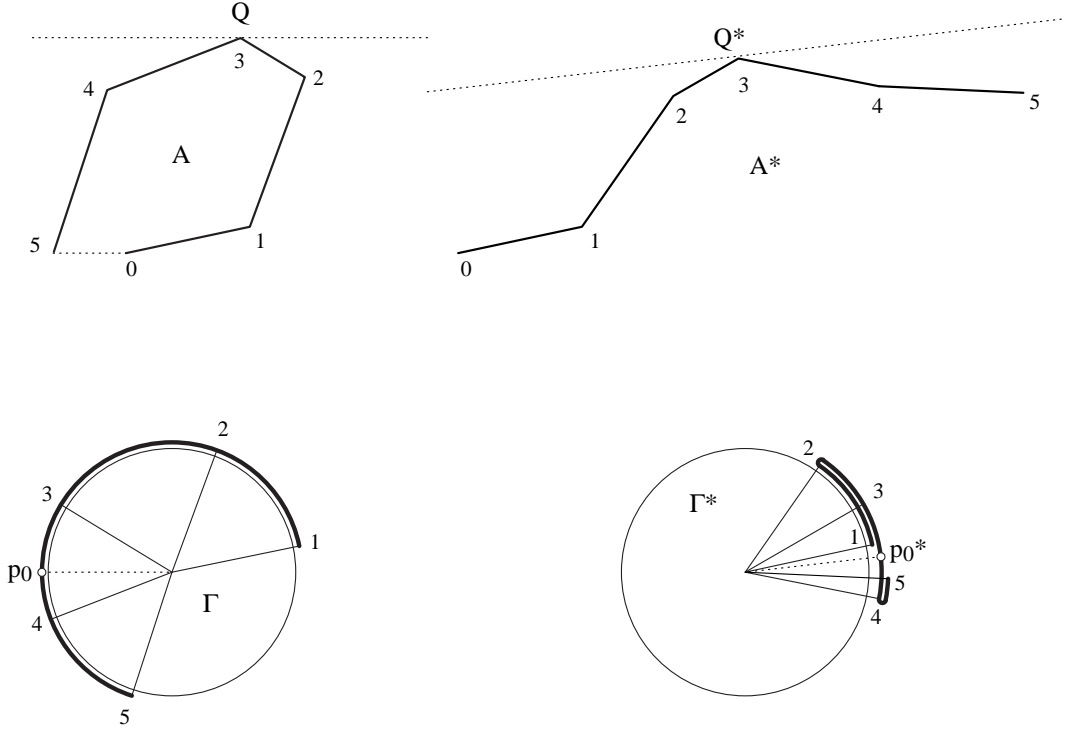}
\caption{$A$ is a convex chain, and $A^*$ a valid
reconfiguration. $\G$ and $\G^*$ are their respective
tangent indicatrixes.}
\figlab{Schur}
\end{figure*}

We will employ two nonnegative measures on an indicatrix.
Let $p_1$ and $p_2$ be any two points of an indicatrix $\G$,
with $p_1$ prior to $p_2$ according to the curve's orientation.
The {\em arc length\/} $\l(p_1,p_2)$ is the length of the
curve $\G$ from $p_1$ to $p_2$.  Note that if the curve
doubles back over itself, this length measure
includes this doubling---there is
no subtraction.
Second, the {\em spherical distance\/} $\s(p_1,p_2)$ is the length of the
shortest path on the sphere from $p_1$ to $p_2$.
For plane curves, when the sphere is a circle, the spherical
distance is the shortest circle arc connecting $p_1$ and $p_2$.
$\s$ is always in $[0,\pi]$.
These definitions imply
\begin{eqnarray}
\s(p_1, p_2) & \le & \l(p_1,p_2) \eqlab{s.l}\\
\s(p^*_1, p^*_2) & \le & \l(p^*_1,p^*_2) \eqlab{s.l*}
\end{eqnarray}
where $p^*_1$ and $p^*_2$ are the points of $\G^*$ corresponding
to $p_1$ and $p_2$ of $\G$.
For example, in Fig.~\figref{Schur}, the arc length
between $1$ and $5$ in $\G$ is approximately $241^\circ$, but
the spherical distance between those points is $119^\circ$.

It should be clear that the arc length $\l$ measures the amount
the tangent turns in absolute value.
The absolute value prevents cancellation when the indicatrix 
doubles back over itself.
Thus $\l(p_1,p_2) = \sum | \a_i | = \sum \a_i$,
where the sum is over all the vertices between,
at which the tangent turn is concentrated.
Note here the absolute value is unnecessary because $A$
is convex and so $\a_i \ge 0$.
Because $A^*=B$ is a valid reconfiguration of $A$, 
Eq.~\eqref{turn.range} holds: $|\b_i| \le a_i$.
So we have
\begin{eqnarray}
\sum | \b_i | & \le & \sum a_i \eqlab{sum.ineq} \\
\l(p^*_1,p^*_2) & \le & \l(p_1,p_2) \eqlab{angle.ineq}
\end{eqnarray}
For example,
in Fig.~\figref{Schur}, the total arc length of $\G^*$ is about
$68^\circ$, considerably less than $\G$'s length of $241^\circ$.

We must address here an issue that does not arise in Chern's proof,
which assumes smoothness of the original curve.
Without a smoothly turning tangent, there may be no unique correspondence
between $p \in \G$ and $p^* \in \G^*$.
For example, let $q = a_i$ be a vertex of $A$.
If we select $p \in \G$ to be the tangent determined by
$a_{i-1} a_i$ and $p^* \in \G^*$ to be the tangent
determined by $a_i a_{i+1}$, then it could be this choice renders
Eq.~\eqref{angle.ineq} false.  For example, suppose $A = A^*$,
$p_1$ is interior to $a_0 a_1$, and $p_2$ is the $p$ just
described.  Then the $\b$-sum in Eq.~\eqref{sum.ineq} includes $\b_i=\a_i$ but
the $\a$-sum excludes $\a_i$, and so the inequality is false.
We repair this problem in the one instance of the proof
that it matters by imagining an intermediate, zero-length
link of the chain $A$ parallel to the relevant tangent, and 
which is present
as well in $A^*$. If the turn angle at that vertex in $A$ is $\a$,
the turn is partitioned to $\a = \a' + \a''$,
and correspondingly in $A^*$ we have $\b = \b' + \b''$ with
$|\b'| \le \a'$ and $|\b''| \le \b''$.
The effect is just as if we had a short link in $A$ aligned
perfectly with the tangent $p$.

Let $Q \in A$ be a point through which a tangent is parallel to
the missing link $a_0 a_n$.
Let $p_0$ be the image of $Q$ on $\G$.
Then, for any other point $p \in \G$, the tangent can turn
at most $\pi$ between $p_0$ and $p$:
\begin{equation}
\l(p_0,p) \le \pi \eqlab{l.pi}
\end{equation}
This means that $p_0$ and $p$ lie within the same semicircle, and
because the convexity of $A$ ensures that $\G$ does not double back
over itself, in this circumstance the spherical distance and the
arc length coincide:
\begin{equation}
\s(p_0,p) = \l(p_0,p) \eqlab{s.eq.l}
\end{equation}
Now we need the corresponding point $p^*_0 \in \G^*$,
and we follow the plan mentioned earlier to obtain a valid
corresponding point.
See Fig.~\figref{Schur}; here $\a_3 = 30^\circ + 22^\circ$,
and so we choose a tangent at $Q^*$ that turns no more than $30^\circ$,
leaving a further turn of no more than $22^\circ$ to the next link.

The issue of carefully choosing a point
$p^*$ of $\G^*$ corresponding to $p \in G$ does not
arise, as we may select $p$ to be directly on one of the
link tangents.
From Eqs.~\eqref{s.l*} and~\eqref{angle.ineq}, using $p_0$ and
$p$
as the two arbitrary points $p_1$ and $p_2$, we have
\begin{eqnarray}
\s(p^*_0, p^*) & \le & \l(p^*_0,p^*) \\
\l(p^*_0,p^*) & \le & \l(p_0,p) \\
\l(p_0,p) & = &  \s(p_0,p)
\end{eqnarray}
that is,
\begin{equation}
\s(p^*_0,p^*) \le \s(p_0,p)  \le \pi \eqlab{s.s}
\end{equation}
The distance $\s(p_0,p)$ may be interpreted as the
angle between the tangent specified by $p$ and the line
determined by $p_0$.
Thus $\cos ( \s(p_0,p) )$ is the projection of a link
of the chain whose tangent is $p$ onto that line.
Thus the distance between the endpoints of the chain $a = |a_0 a_n|$
may be computed as
\begin{equation}
a = \int_0^L \cos ( \s(p_0,p) ) ds
\end{equation}
where $p$ varies with parameter $s$ over the entire length $L$ of
chain $A$.
For a polygonal chain, this reduces to a sum
\begin{equation}
a = \sum_{i=0}^{n-1} \ell_i \cos( \t_i ) \eqlab{a}
\end{equation}
where $\t_i$ is the angle of link $i$ from the line
through $Q$, i.e., $\s(p_0,p)$.
This expression can be viewed as computing the 
(horizontal) $x$-coordinates
of $a_0$ and $a_n$ working in both directions from $Q$.

Now we look at the corresponding expression for $\G^*$:
\begin{equation}
\int_0^L \cos ( \s(p^*_0,p^*) ) ds
\end{equation} 
This is again a projection of the curve $A^*$, and therefore of
its missing edge, but onto the line determined by $p^*_0$.
This line has no particular geometric significance;
in particular, it is not necessarily parallel to the chord between the
endpoints of $A^*$.
However, because a projection is never longer than the original,
it provides a lower bound on that chord length $a^*=|b_0 b_n|$:
\begin{equation}
a^* \ge \sum_{i=0}^{n-1} \ell_i \cos( \t^*_i ) \eqlab{a*}
\end{equation}
where $\t^*_i$ is the angle of link $i$ of $A^*$ from the
line determined by $p^*_0$.
Finally, we observe from Eq.~\eqref{s.s} that $\t^*_i \le \t_i \le \pi$,
and because the cosine function is monotone decreasing over $[0,\pi]$,
we have
\begin{equation}
\cos( \t^*_i ) \ge \cos (\t_i)
\end{equation}
With Eqs.~\eqref{a} and~\eqref{a*} this finally implies that 
$a^* \ge a$, i.e., the distance between the endpoints has increased
(or stayed the same).
\end{pf}

\noindent
As Connelly remarks~\cite{c-re-82},
this proof can be viewed as a variant of Zaremba's ``shadow'' proof of
Cauchy's arm lemma~\cite{sz-clccp-67}.

\section{Noncrossing of Straightened Curve}
\seclab{Noncrossing}
Define a polygonal chain to be {\em simple\/} if nonadjacent
segments are disjoint, and adjacent segments intersect only at their
single, shared endpoint.
By our nondegeneracy requirement, convex chains are simple.
In particular, any opening of a convex chain via Cauchy's
arm lemma (Theorem~\theoref{cauchy}) remains simple
because it remains convex.
We now establish a parallel result for the generalized
straightening of Theorem~\theoref{cgen}.
We generalize slightly to permit the convex chain to start
with the hand at the shoulder.

\begin{theorem}
If $A = (a_0,\ldots,a_n) = \C_L(\a)$ 
is a closed convex chain with $n$ fixed edge lengths $L$ and
turn angles $\a$, 
closed in the sense that $a_n = a_0$,
then any valid reconfiguration to
$B = \C_L(\b)$ 
is a simple polygonal chain.
\theolab{noncrossing}
\end{theorem}
\begin{pf}
Suppose to the contrary that $B$ is nonsimple.
Let $q_2$ be the first point of $B$, measured by distance
along the chain from the shoulder $b_0$, that coincides
with an earlier point $q_1 \in B$.
Thus $q_1$ and $q_2$ represent the same point of the plane,
but different points along $B$.
See Fig.~\figref{cross}.
Because $B$ is nonsimple, these ``first touching points''
exist,\footnote{
	The proof works for any self-intersection point.
	We only choose the first for definiteness.}
and we do not have both $q_1 = b_0$ and $q_2 = b_n$
(because that would make $B$ a simple, closed chain).
Let $p_1$ and $p_2$ be the points of $A$ corresponding to $q_1$ and $q_2$.
\begin{figure}[htbp]
\centering
\includegraphics[width=0.8\linewidth]{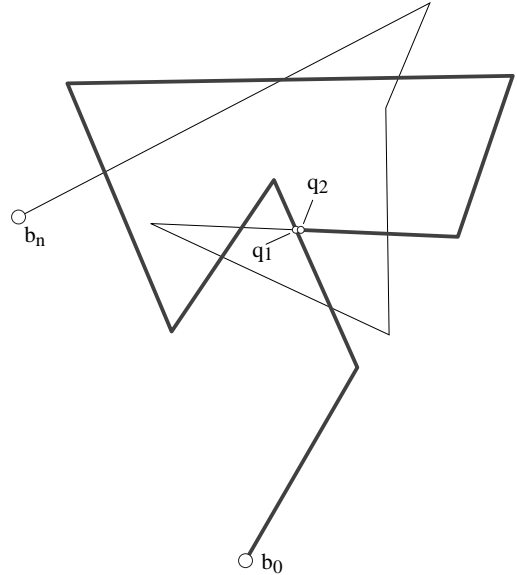}
\caption{Violation of Theorem~\theoref{cgen}. 
$q_1=q_2$ is the first point of self-contact; the initial
portion of $B$, up to $q_2$, is highlighted.}
\figlab{cross}
\end{figure}
Corollary~\corref{all.points} guarantees
that
$|q_1 q_2| \ge |p_1 p_2|$.
But $|q_1 q_2| = 0$, and because the $q$'s do not coincide
with the original hand and shoulder, $|p_1 p_2| >0$.
This contradiction establishes the claim.
\end{pf}

\noindent
One could alternatively prove this theorem 
by induction on the length of the chain, 
showing that in a continuous motion to $B$,
the first violation of simplicity is either impossible by
the induction hypothesis, or directly contradicts Theorem~\theoref{cgen}.

\begin{cor}
A valid reconfiguration of an open convex chain
remains simple.
\end{cor}
\begin{pf}
Theorem~\theoref{noncrossing} guarantees that even the
final missing edge between $a_n$ and $a_0$ is not crossed,
so the corollary is obtained by simply ignoring that last
edge.
\end{pf}

\section{Application to Curve Development}
\seclab{slice}

A curve $\G$ on the surface of a convex body may be ``developed''
on a plane by rolling the convex body on the plane without slippage
so that the curve is always the point of contact.
Here we will only consider polygonal curves on the surface of
convex polyhedra (polytopes).
An earlier result is that a closed convex polygonal curve on a polytope,
i.e., one whose turns are all leftward on the surface,
develops to a simple path~\cite{os-odcc3p-89}.
Here we prove that particular (nonconvex) curves also develop
without self-intersection:
{\em slice\/} curves, those that are the intersection of a polytope
with a plane; see
Fig.~\figref{slice}.
\begin{figure}[htbp]
\centering
\includegraphics[width=0.9\linewidth]{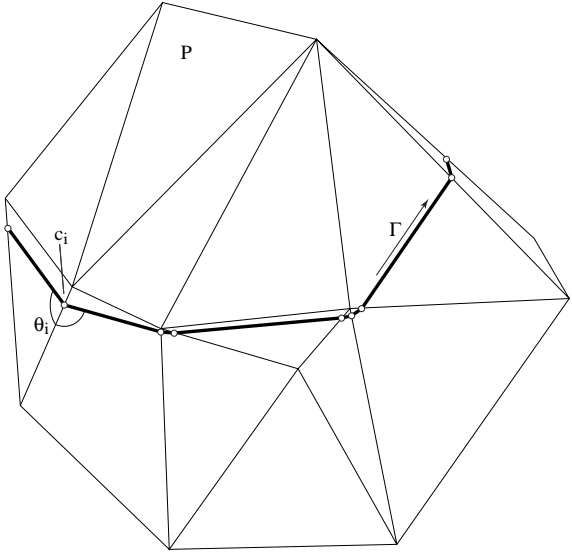}
\caption{$\G$ is the intersection of a plane (not shown) with polytope $P$.}
\figlab{slice}
\end{figure}

Orient $\G$ to be counterclockwise from
above.
Let $c_0,c_1,\ldots,c_n$ be the {\em corners\/} of $\G$, the
points at which $\G$ crosses a polytope edge
with
a dihedral angle different from $\pi$, or meets a polytope vertex.
Define the {\em right surface angle\/} $\t(p)$ at a point
$p \in \G$ to be the total incident face angle at $p$ to the
right of the directed curve $\G$ at $p$.
Only at a corner $c_i$ of $\G$ is the right surface angle $\t_i$
different from $\pi$.  Note that $\t_i$ could be greater or less
than $\pi$, i.e., the slice curve could turn right or left
on the surface.

Define the {\em right development\/} of $\G$ to be a
planar drawing of the polygonal chain $\G$ as
the chain $B = (b_0,b_1,\ldots,b_n)$ with the same link lengths,
$|b_i b_{i+1}| = |c_i c_{i+1}|$ for $i=0,\ldots,n-1$,
and with exterior angle $\t_i$ to the right of $b_i$ 
the same as the surface angle
to the right of $\G$ at $c_i$ on $P$, for all $i=1,\ldots,n-1$.
Define {\em left development\/} similarly.  Note that if $\G$
avoids all polytope vertices, then there is no difference between
the left and right development of $\G$, for the sum of the
right and right surface angles at any point is always $2 \pi$.
Define the {\em development\/} of $\G$ to be the right development of $\G$.

\begin{theorem}
Let $\G = P \cap \Pi$ be a 
closed curve on the surface of a polytope $P$
that is the intersection of $P$ with a plane $\Pi$.
Then $\G$ develops on a plane to a simple (noncrossing) polygonal curve.
\theolab{slice}
\end{theorem}
\begin{pf}
We first dispense with the degenerate intersections,
where there is zero volume of $P$ to one side of $\Pi$.
Then $\Pi$ must intersect $P$ in either a face, an edge,
or a vertex.  In all cases, $\G$ develops as is, and there is
nothing to prove.  Henceforth we assume that the slice is
nondegenerate.

Let $Q$ be the convex polygon in plane $\Pi$ bound by $\G$.
Let $\p_i \in (0,\pi)$ be the internal convex angle of $Q$ at $c_i$.
Our aim is to prove that these internal angles are related
to the right surface angles $\t_i$ as follows:
\begin{equation}
\p_i \le \t_i \le 2\pi - \p_i
\eqlab{angles}
\end{equation}
First note that, by our nondegeneracy assumption,
the intersection of $P$ with the halfspace
below (and including) $\Pi$ is a polytope; call it $P_0$.
$P_0$ has $Q$ as a top face, and the corners of $\G$ as
vertices.  The total angle incident to vertex $c_i$ of
$P_0$ is $\p_i + \t_i$, because $P_0$ includes the entire
right surface angle at $c_i$.
Because $P_0$ is a polytope, this sum must be at most $2 \pi$,
and from 
$\p_i + \t_i  \le 2 \pi$
the right hand inequality of Eq.~\eqref{angles} follows.

Repeating the argument to the
other side, let $P_1$ be the intersection of $P$ with the halfspace
above $\Pi$.
Because the left surface angle at $c_i$ on $P$ is no more
than $2\pi - \t_i$ (with equality if $c_i$ is not a vertex
of $P$), then the total angle incident to vertex $c_i$ of
$P_1$ is no more than  $\p_i + 2\pi - \t_i$.
Because $P_1$ is a polytope, this sum must be at most $2 \pi$,
and from 
$\p_i + 2\pi - \t_i\le 2 \pi$
the left hand inequality of Eq.~\eqref{angles} follows.

Let $A = (a_0,a_1,\ldots,a_n)$ be a polygonal chain
representing convex polygon $Q$, with $a_i$ corresponding to $c_i$.
The turn angle $\a_i$ at $a_i$ is $\a_i = \pi - \p_i$,
i.e.,
\begin{equation}
\p_i = \pi - \a_i
\eqlab{phi}
\end{equation}
Let $B = (b_0,b_1,\ldots,b_n)$ be the development of $\G$ in
the plane, again with $b_i$ corresponding to $c_i$.
The turn angle $\b_i$ at $b_i$ is determined by the right
surface angle:  $\b_i = \t_i - \pi$, i.e.,
\begin{equation}
\t_i = \pi + \b_i
\eqlab{theta}
\end{equation}
Substituting Eqs.~\eqref{phi} and~\eqref{theta} into Eq.~\eqref{angles}
yields
$$
\begin{array}{ccccc}
\pi - \a_i & \le & \pi + \b_i & \le & 2 \pi - (\pi - \a_i) \\
-\a_i & \le & \b_i & \le & \a_i 
\end{array}
$$
i.e., $\b_i \in [-\a_i, \a_i]$.
Thus we see that $B$ is a valid reconfiguration of $A$, and
Theorem~\theoref{noncrossing} applies and establishes the claim
that it is simple.
\end{pf}

\noindent
The examples in Fig.~\figref{examples} can all be viewed as
developments of slice curves.

Because Schur's Theorem generalizes to smooth curves,
Theorem~\theoref{slice} should generalize to
slice curves for any convex body $B$.

\paragraph{Acknowledgements.}
I thank
Erik Demaine,
Martin Demaine,
Anna Lubiw, and
Godfried Toussaint
for a 
clarifying discussion at a critical juncture of the proof
of Theorem~\theoref{cgen}.
In particular, Anna Lubiw suggested the proof in Section~\secref{proof.2} 
and simplified the proof of Theorem~\theoref{noncrossing}.
I thank Veronica Morales for writing the code that produced
Fig.~\figref{examples}.
I thank Michael Albertson for a technical suggestion.
Finally, I thank Raghavan Dhandapani for detecting two errors
in the proofs.

\bibliographystyle{alpha}
\bibliography{arm}

\begin{thebibliography}{HJW84}

\bibitem[AZ98]{az-pftb-98}
M.~Aigner and G.~M. Ziegler.
\newblock {\em Proofs from {THE} {BOOK}}.
\newblock Springer-Verlag, Berlin, 1998.

\bibitem[Che89]{c-cses-89}
S.~S. Chern.
\newblock Curves and surfaces in {E}ucidean space.
\newblock In S.~S. Chern, editor, {\em Global Differential Geometry}, volume~27
  of {\em Studies in Mathmatics}, pages 99--139. Math. Assoc. Amer., 1989.

\bibitem[Con82]{c-re-82}
R.~Connelly.
\newblock Rigidity and energy.
\newblock {\em Invent. Math.}, 66:11--33, 1982.

\bibitem[Cro97]{c-p-97}
P.~Cromwell.
\newblock {\em Polyhedra}.
\newblock Cambridge University Press, 1997.

\bibitem[Gug63]{g-dg-63}
H.~W. Guggenheimer.
\newblock {\em Differential Geometry}.
\newblock McGraw-Hill, 1963.

\bibitem[HJW84]{hjw-mp2dl-84}
J.~E. Hopcroft, D.~A. Joseph, and S.~H. Whitesides.
\newblock Movement problems for $2$-dimensional linkages.
\newblock {\em SIAM J. Comput.}, 13:610--629, 1984.

\bibitem[O'R98]{o-cgc-98}
J.~O'Rourke.
\newblock {\em Computational Geometry in {C} (Second Edition)}.
\newblock Cambridge University Press, 1998.

\bibitem[OS89]{os-odcc3p-89}
J.~O'Rourke and C.~Schevon.
\newblock On the development of closed convex curves on 3-polytopes.
\newblock {\em J. Geom.}, 13:152--157, 1989.

\bibitem[Sch21]{s-usedkkk-21}
A.~Schur.
\newblock {\"U}ber die {S}chwarzche {E}xtremaleigenschaft des {K}reises unter
  den {K}urven konstantes {K}r{\"u}mmung.
\newblock {\em Math. Ann.}, 83:143--148, 1921.

\bibitem[Sin97]{s-gpf-97}
D.~Singer.
\newblock {\em Geometry: Plane and Fancy}.
\newblock Springer-Verlag, Berlin, 1997.

\bibitem[SZ67]{sz-clccp-67}
I.~J. Schoenberg and S.~K. Zaremba.
\newblock On {Cauchy}'s lemma concerning convex polygons.
\newblock {\em Canad. J. Math.}, 19:1062--1077, 1967.

\end{thebibliography}

\onecolumn
\begin{figure}[htbp]
\centering
\includegraphics[width=0.75\linewidth]{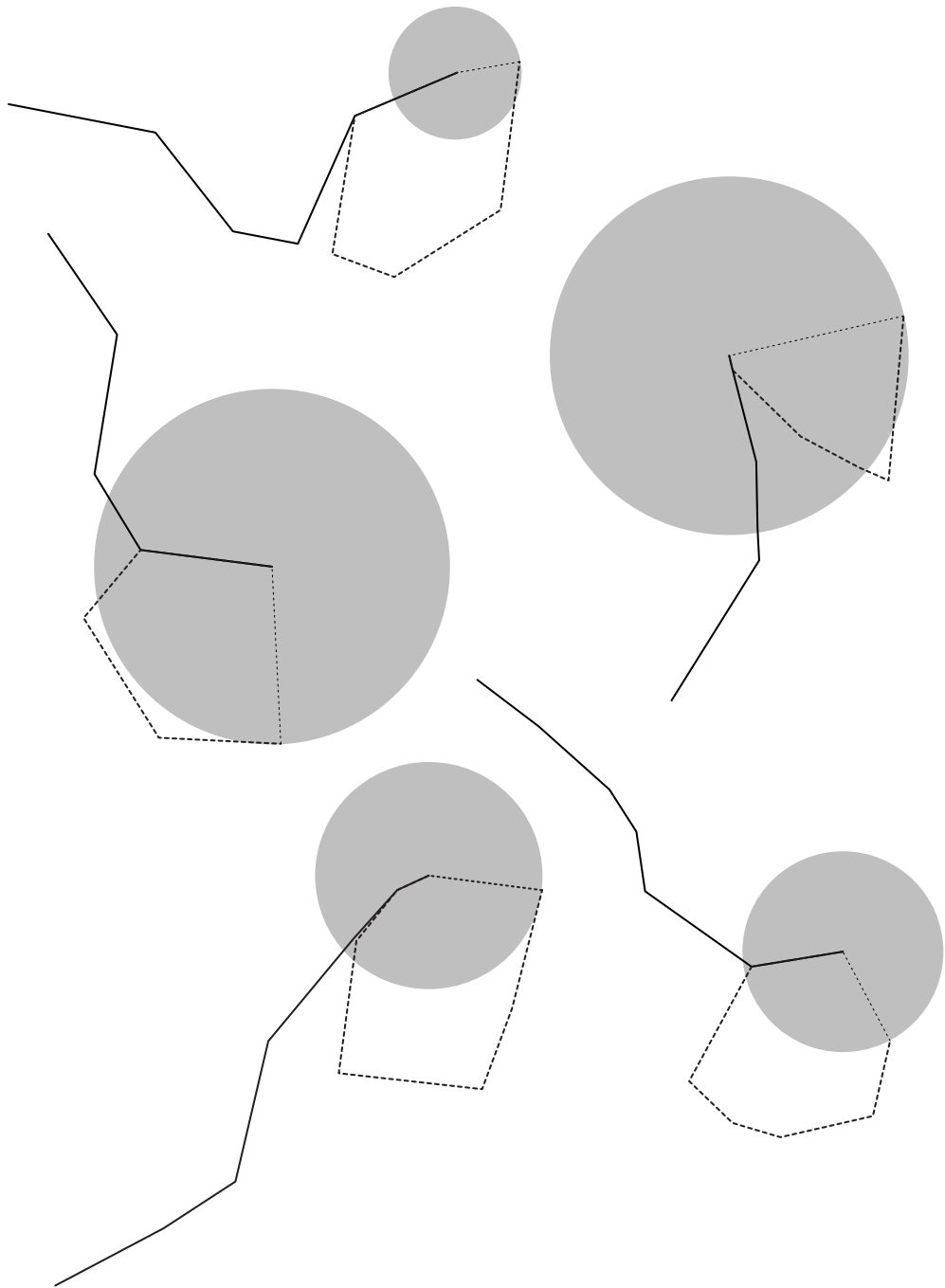}
\caption{Examples of valid reconfigurings
of convex chain $A$ (dashed) to $B$ (solid).  In each case,
the forbidden shoulder disk is shown.}
\figlab{examples}
\end{figure}

\end{document}